\documentclass[journal,hidelinks]{IEEEtran}

\ifCLASSINFOpdf
\else
\usepackage[dvips]{graphicx}
\fi
\usepackage{url}

\usepackage{graphicx} 
\usepackage{booktabs} 
\usepackage{mathrsfs}
\usepackage{amsmath}
\usepackage{amssymb}
\usepackage{lipsum}
\usepackage{mathtools}
\usepackage{cite}
\usepackage[acronym,shortcuts]{glossaries}
\usepackage{lipsum}
\usepackage{xcolor}

\usepackage{comment}

\usepackage{algorithm, algpseudocode}
\usepackage{algcompatible}
\usepackage{subfigure}

\usepackage{orcidlink}
\usepackage{bm}

\newcommand{\trans}[0]{^{\mathsf{T}}}

\newcommand{\herm}[0]{^{\mathsf{H}}}

\newacronym{FA}{FA}{flexible antenna}
\newacronym{REMS}{REMS}{reconfigurable electromagnetic structure}
\newacronym{DSA}{DSA}{dynamic scattering array}
\newacronym{RPE}{RPE}{radar parameter estimation}
\newacronym{OTFS}{OTFS}{orthogonal time frequency space}
\newacronym{AFDM}{AFDM}{affine frequency division multiplexing}
\newacronym{MIMO}{MIMO}{multiple-input multiple-output}
\newacronym{SISO}{SISO}{single-input single-output}
\newacronym{ISAC}{ISAC}{integrated sensing and communications}
\newacronym{ICC}{ICC}{integrated communications and computing}
\newacronym{DT}{DT}{digital twins}
\newacronym{XR}{XR}{virtual and augmented reality}
\newacronym{3D}{3D}{three-dimensional}
\newacronym{D2D}{D2D}{device-to-device}
\newacronym{2D}{2D}{two-dimensional}
\newacronym{1D}{1D}{one-dimensional}
\newacronym{AI}{AI}{artificial intelligence}
\newacronym{RX}{RX}{receiver}
\newacronym{TX}{TX}{transmitter}
\newacronym{BF}{BF}{beamforming}
\newacronym{mmWave}{mmWave}{millimeter-wave}
\newacronym{CAV}{CAV}{connected autonomous vehicles}
\newacronym{SotA}{SotA}{state-of-the-art}
\newacronym{ULA}{ULA}{uniform linear array}
\newacronym{QAM}{QAM}{quadrature amplitude modulation}
\newacronym{ISFFT}{ISFFT}{inverse symplectic finite Fourier transform}
\newacronym{SFFT}{SFFT}{symplectic finite Fourier transform}
\newacronym{AWGN}{AWGN}{additive white Gaussian noise}
\newacronym{OFDM}{OFDM}{orthogonal frequency division multiplexing}
\newacronym{OCDM}{OCDM}{orthogonal chirp division multiplexing}
\newacronym{BS}{BS}{base station}
\newacronym{UE}{UE}{user equipment}
\newacronym{DFT}{DFT}{discrete Fourier transform}
\newacronym{IDFT}{IDFT}{inverse discrete Fourier transform}
\newacronym{TD}{TD}{time-domain}
\newacronym{wlg}{wlg}{without loss of generality}
\newacronym{CP}{CP}{cyclic prefix}
\newacronym{DAFT}{DAFT}{discrete affine Fourier transform}
\newacronym{IDAFT}{IDAFT}{inverse discrete affine Fourier transform}
\newacronym{CPP}{CPP}{\textit{chirp-periodic} prefix}
\newacronym{IDZT}{IDZT}{inverse discrete Zak transform}
\newacronym{DZT}{DZT}{discrete Zak transform}
\newacronym{ICI}{ICI}{inter-carrier interference}
\newacronym{BER}{BER}{bit error rate}
\newacronym{DoF}{DoF}{degrees-of-freedom}
\newacronym{FD}{FD}{full-duplex}
\newacronym{SIMO}{SIMO}{single-input multiple-output}
\newacronym{MISO}{MISO}{multiple-input single-output}
\newacronym{AoD}{AoD}{angle-of-departure}
\newacronym{AoA}{AoA}{angle-of-arrival}
\newacronym{RF}{RF}{radio frequency}
\newacronym{SIM}{SIM}{stacked intelligent metasurfaces}
\newacronym{FPGA}{FPGA}{field programmable gate array}
\newacronym{UPA}{UPA}{uniform planar array}
\newacronym{CC}{CC}{communication-centric}
\newacronym{I/O}{I/O}{input-output}
\newacronym{iid}{i.i.d.}{independent and identically distributed}
\newacronym{IoT}{IoT}{internet of things}
\newacronym{V2X}{V2X}{vehicle-to-everything}
\newacronym{NTN}{NTN}{non-terrestrial network}
\newacronym{LEO}{LEO}{low-earth orbit}
\newacronym{THz}{THz}{terahertz}
\newacronym{EM}{EM}{expectation maximization}
\newacronym{RIS}{RIS}{reconfigurable intelligent surface}
\newacronym{DoA}{DoA}{direction-of-arrival}
\newacronym{DD}{DD}{doubly-dispersive}
\newacronym{ODDM}{ODDM}{orthogonal delay-Doppler division multiplexing}
\newacronym{LoS}{LoS}{line-of-sight}
\newacronym{NLoS}{NLoS}{non-line-of-sight}
\newacronym{5G}{5G}{fifth generation}
\newacronym{6G}{6G}{sixth generation}
\newacronym{MPDD}{MPDD}{metasurfaces-parametrized doubly-dispersive}
\newacronym{GaBP}{GaBP}{Gaussian belief propagation}
\newacronym{MSE}{MSE}{mean-squared-error}
\newacronym{sIC}{soft IC}{soft interference cancellation}
\newacronym{soft RG}{soft RG}{soft replica generation}
\newacronym{BG}{BG}{belief generation}
\newacronym{SGA}{SGA}{scalar Gaussian approximation}
\newacronym{CLT}{CLT}{central limit theorem}
\newacronym{PDF}{PDF}{probability density function}
\newacronym{QPSK}{QPSK}{quadrature phase-shift keying}
\newacronym{LMMSE}{LMMSE}{linear minimum mean square error}
\newacronym{SNR}{SNR}{signal-to-noise ratio}
\newacronym{LTV}{LTV}{linear time-variant}
\newacronym{TVIRF}{TVIRF}{time-variant impulse response function}
\newacronym{FT}{FT}{Fourier transform}
\newacronym{TVTF}{TVTF}{time-variant transfer function}
\newacronym{DVIRF}{DVIRF}{Doppler-variant impulse response function}
\newacronym{CSI}{CSI}{channel state information}
\newacronym{CRLB}{CRLB}{Cram{\`e}r-Rao lower bound}
\newacronym{BCRLB}{BCRLB}{Bayesian Cram{\`e}r-Rao lower bound}
\newacronym{BBI}{BBI}{Bayesian bilinear inference}
\newacronym{ML}{ML}{maximum likelihood}
\newacronym{MUSIC}{MUSIC}{multiple signal classification}
\newacronym{MU}{MU}{multi-user}
\newacronym{ROOT-MUSIC}{ROOT-MUSIC}{ROOT multiple signal classification}
\newacronym{JCAS}{JCAS}{joint communication and sensing}
\newacronym{JCR}{JCR}{joint communications and radar}
\newacronym{ROI}{ROI}{region of interest}
\newacronym{MF}{MF}{matched-filter}
\newacronym{ISI}{ISI}{inter-symbol interference}
\newacronym{RMSE}{RMSE}{root mean square error}
\newacronym{ESPRIT}{ESPRIT}{estimation of signal parameters via rotational invariant techniques}
\newacronym{JCDE}{JCDE}{joint channel and data estimation}
\newacronym{PDA}{PDA}{probabilistic data association}
\newacronym{PMF}{PMF}{probability mass function}
\newacronym{PBiGaBP}{PBiGaBP}{parametric bilinear Gaussian belief propagation}
\newacronym{PBiGAMP}{PBiGAMP}{parametric bilinear generalized approximate message passing}
\newacronym{GAMP}{GAMP}{generalized approximate message passing}
\newacronym{EP}{EP}{expectation propagation}
\newacronym{DAF}{DAF}{discrete affine Fourier}
\newacronym{P/S}{P/S}{parallel-to-serial}
\newacronym{S/P}{S/P}{serial-to-parallel}
\newacronym{SBL}{SBL}{sparse Bayesian learning}
\newacronym{MPA}{MPA}{message passing algorithms}
\newacronym{VGA}{VGA}{vector Gaussian approximation}
\newacronym{SIC}{SIC}{self-interference cancellation}
\newacronym{NMSE}{NMSE}{normalized mean square error}
\newacronym{KL}{KL}{Kullback-Leibler}
\newacronym{i.i.d.}{i.i.d.}{independent and identically distributed}

\begin{document}

\title{Parametrized Stacked Intelligent Metasurfaces for Bistatic Integrated Sensing and Communications}

\author{Kuranage Roche Rayan Ranasinghe\textsuperscript{\orcidlink{0000-0002-6834-8877}},
Iv{\'a}n Alexander Morales Sandoval\textsuperscript{\orcidlink{0000-0002-8601-5451}},~\IEEEmembership{Graduate Student Members,~IEEE,}\\
Giuseppe Thadeu Freitas de Abreu\textsuperscript{\orcidlink{0000-0002-5018-8174}} and
George C. Alexandropoulos\textsuperscript{\orcidlink{0000-0002-6587-1371}},~\IEEEmembership{Senior Members,~IEEE}
\thanks{K. R. R. Ranasinghe, I. A. M. Sandoval and G. T. F. de Abreu are with the School of Computer Science and Engineering, Constructor University (previously Jacobs University Bremen), Campus Ring 1, 28759 Bremen, Germany (emails: \{kranasinghe, imorales, gabreu\}@constructor.university).} 
\thanks{G. C. Alexandropoulos is with the Department of Informatics and Telecommunications, National and Kapodistrian University of Athens, 15784 Athens, Greece and with the Department of Electrical and Computer Engineering, University of Illinois Chicago, Chicago, IL 60601, USA (e-mail: alexandg@di.uoa.gr).
}\vspace{-3ex}}

\maketitle

\begin{abstract}
We consider \acp{SIM} as a tool to improve the performance of bistatic \ac{ISAC} schemes.
To that end, we optimize the \acp{SIM} and design a \ac{RPE} scheme aimed at enhancing radar sensing capabilities as well as communication performance under \ac{ISAC}-enabling waveforms known to perform well in \ac{DD} channels.
The \ac{SIM} optimization is done via a min-max problem formulation solved via steepest ascent with closed-form gradients, while the \ac{RPE} is carried out via a compressed sensing-based \ac{PDA} algorithm.
Our numerical results indicate that the design of waveforms suitable to mitigating the effects of \ac{DD} channels is significantly impacted by the emerging \ac{SIM} technology.
\end{abstract}


\IEEEpeerreviewmaketitle

\glsresetall

\vspace{-3ex}
\section{Introduction}

%
%
\IEEEPARstart{T}{he} \ac{6G} of wireless communications systems is expected to bring about a host of new functionalities such as support to and from \ac{AI} \cite{CelikOJCS2024}, \ac{D2D} connectivity \cite{GismallaD2D2022}, incorporation of \acp{NTN} \cite{AzariCST2022}, and \ac{ISAC} as well as \ac{ICC} \cite{QiaoTCOM2022, HyeonTWC2024, Ranasinghe2024_ICC}, which are essential to enable novel applications such as \ac{CAV} \cite{JianhuaNet2021}, \ac{DT} \cite{Guo_WN2024}, \ac{XR} \cite{StafidasAccess2024} and more.
These applications require either robustness against high mobility scenarios \cite{LiCOMMST2022} or high data rates combined with extremely low latencies, which explains recent trends for \ac{6G} to move to higher frequency bands and, in turn, give rise to the necessity of using \ac{DD} channel models \cite{LiuTIT2004, Bliss_Govindasamy_2013, ZhangTVT2017}.

%
%
The \ac{DD} channel model captures some of the most significant effects undergone by a time-domain transmit signal propagating through a channel with fast mobility and/or at high frequencies, and it has been shown that, if well-accounted for, its properties can be leveraged to improve multiple-access \cite{BuzziDD2010}, reap diversity \cite{SurabhiTWC2019}, and (more recently) enable the estimation of \ac{RPE} in support of \ac{ISAC} \cite{GaudioTWC2020,KuranageTWC2024, RanasingheWCNC2025}. 
The latter \ac{ISAC}-supporting feature of \ac{DD} channels, in particular, has motivated intense investigation on the performance of \ac{OFDM} in such channels \cite{HeTSP2011}, as well as the design of new waveforms capable of simultaneously combating undesired effects, and extracting the information made available by \ac{DD} channels \cite{WeiTIoT2023, Tagliaferri_TWC_2023, Chakravarthi2024}, with two prominent examples being \ac{OTFS} modulation and \ac{AFDM} \cite{Mohammed_BITS_2022, Bemani_TWC_2023, Rou_SPM_2024}.

%
%
Somewhat disconnected to such a quest for \ac{DD}-suitable waveforms, a significant amount of work has been recently done on the design and modeling of \acp{REMS} \cite{DardariJSAC2024, Studer2024}, which include \acp{RIS} \cite{LiaskosCommMag2018}, \acp{SIM} \cite{AnJSAC2023}, \acp{DSA} \cite{DardariDSA2024}, \acp{FA} \cite{KiatFA6g2024} and others.
In fact, to the best of our knowledge, the work in \cite{RanasingheDDSIM2025} is the only contribution\footnote{The only other related work we are aware of is \cite{ChaoITJ2023} where, focusing on a space-to-ground setting, it was shown in that an \ac{OTFS} system aided by a \ac{RIS} is significantly enhanced when the \ac{RIS} is optimized directly over the delay-Doppler domain, as opposed to the time-frequency.
The model described thereby is, however, limited to the \ac{OTFS} waveform and rather simple, incorporating only one \ac{RIS} and a single antenna assumed at both the \ac{TX} and \ac{RX}.} thus far which sought to provide a thorough \ac{DD} channel model incorporating \ac{MIMO}, \acp{RIS} and \acp{SIM}, and which can be applied to multiple \ac{ISAC}-enabling waveforms.

%
%
To close the aforementioned gap, we developed in \cite{RanasingheDDSIM2025} a complete \acp{REMS}-inclusive \ac{DD}-\ac{MIMO} channel model, incorporating \acp{SIM} at both the \ac{TX} and \ac{RX}, as well as an arbitrary number of \acp{RIS} in the environment.
The focus of \cite{RanasingheDDSIM2025} was, however, the detailed description of the model\footnote{For the sake of illustration, a very brief discussion was also offered in \cite{RanasingheDDSIM2025} on how \acp{SIM} can be optimized to improve communication performance under a \acp{REMS}-\ac{DD} channel in the particular case of a \ac{SISO} system employing data detection via \ac{GaBP}.}, whereas in this paper the application of the proposed model to an \ac{ISAC} setting is addressed.
In particular, we propose a novel bistatic \ac{ISAC} transceiver architecture leveraging the \ac{PDA} algorithm capable of improving both communication and radar sensing performances, with the \acp{SIM} being optimized to maximize the weakest channel path.
We highlight that thanks to the flexibility of the \acp{REMS}-\ac{DD}-\ac{MIMO} model of \cite{RanasingheDDSIM2025}, the performance of the proposed \ac{ISAC} scheme in \ac{DD} channels can be seamlessly assessed under various waveforms, with simulation results shown for \ac{OFDM}, \ac{OTFS} and \ac{AFDM} signaling, which indicate a significant gain in both functionalities among all these waveforms.

\noindent \textit{Notation:} The following notation will be persistently used hereafter.
All scalars are represented by normal upper or lowercase letters, while column vectors and matrices are denoted by bold lowercase and uppercase letters, respectively.
The diagonal matrix constructed from a vector $\mathbf{a}$ is denoted by diag($\mathbf{a}$), while $\mathbf{A}\trans$, $\mathbf{A}\herm$, $\mathbf{A}^{1/2}$, and $[\mathbf{A}]_{i,j}$ denote the transpose, Hermitian, square root and the $(i,j)$-th element of a matrix $\mathbf{A}$, respectively.
The convolution and Kronecker product are respectively denoted by $*$ and $\otimes$, while $\mathbf{I}_N$ and $\mathbf{F}_N$ represent the $N\times N$ identity and the normalized $N$-point \ac{DFT} matrices, respectively.
The sinc function is expressed as $\text{sinc}(a) \triangleq \frac{\sin(\pi a)}{\pi a}$, and $\jmath\triangleq\sqrt{-1}$ denotes the elementary complex number.

\vspace{-1ex}
\section{System Setup and Signal Model}
\label{sec:system_model}

\vspace{-0.5ex}
\subsection{ISAC System Setup and MPDD Channel Model}
\label{sec:system_and_channel_model}
\vspace{-0.5ex}

\begin{figure}[H]
\centering
\includegraphics[width=\columnwidth]{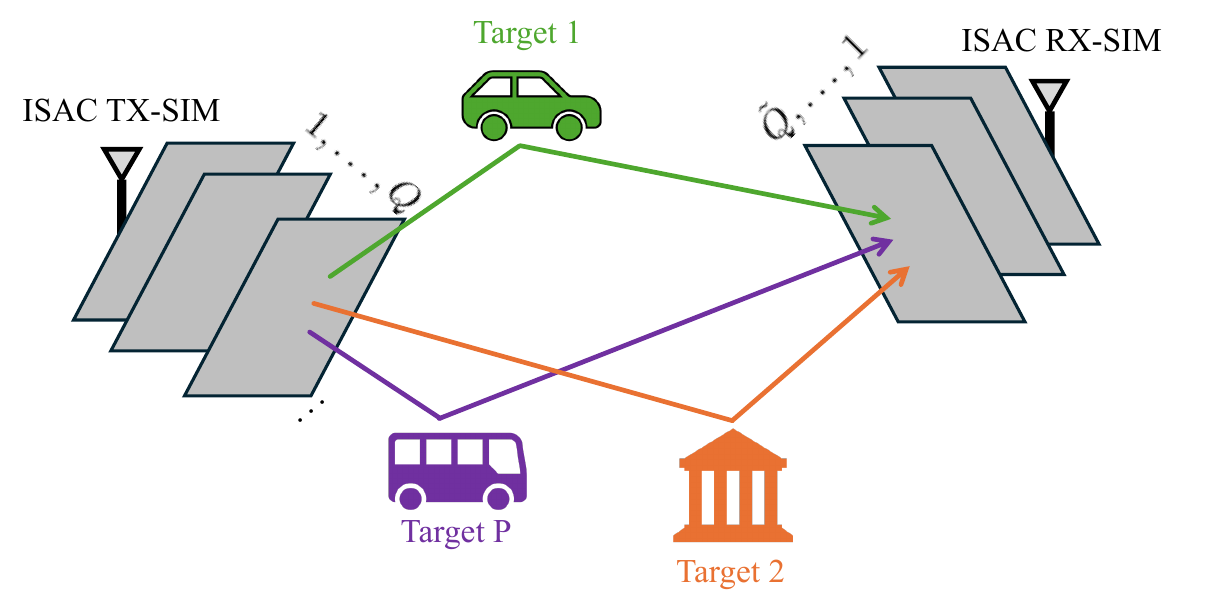}
\vspace{-5ex}
\caption{A bistatic \ac{ISAC} system for high-mobility scenarios, comprised of two \acp{SIM}, one placed very close to the \ac{TX} and the other very close to the \ac{RX}.}
\label{fig:system_model_SIM}
\end{figure}

Consider the system illustrated in Fig. \ref{fig:system_model_SIM}, where a \ac{TX} \ac{BS} downlinks information to various \ac{UE}, while another \ac{RX} \ac{BS} seeks to perform \ac{RPE} -- $i.e.$, estimate the distances and velocities -- of the latter with full knowledge of the \ac{TX} signal\footnotemark, altogether yielding a bistatic \ac{ISAC} scenario\footnotemark.







It is assumed that each \acp{BS} is equipped with a single antenna enhanced by a \ac{SIM}, but that there is no \ac{RIS} in the environment, such that the channel is a metasurfaces-parametrized channel, with transfer functions of the $Q$-layer \ac{TX}-\ac{SIM} with $M$ meta-atoms per layer, and the $\tilde{Q}$-layer \ac{RX}-\ac{SIM} with $\tilde{M}$ meta-atoms per layer, respectively given by
\vspace{-1ex}
\begin{equation}
\bm{v} \triangleq \prod_{q=1}^Q \bm{\Psi}_{Q-q+1} \bm{\Gamma}_{Q-q+1} \in \mathbb{C}^{M \times 1},
\label{eq:transmit_SIM_full}
\vspace{-1ex}
\end{equation}
and 
\vspace{-1ex}
\begin{equation}
\bm{u} \triangleq \prod_{\tilde{q}=1}^{\tilde{Q}} \bm{\Xi}_{\tilde{q}} \bm{\Delta}_{\tilde{q}} \in \mathbb{C}^{1 \times \tilde{M}},
\label{eq:receive_SIM_full}
\vspace{-1ex}
\end{equation}
where the matrices $\bm{\Gamma}_{q}\in \mathbb{C}^{M \times 1}$ and $\bm{\Xi}_{\tilde{q}} \in \mathbb{C}^{1 \times \tilde{M}}$ are typically modeled via the Rayleigh-Sommerfeld diffraction theorem \cite{AnJSAC2023}, and $\bm{\Psi}_q  \triangleq \text{diag}\big( \!\big[ e^{\jmath\zeta^{q}_{1}}, \dots, e^{\jmath\zeta^{q}_{M}} \big]\! \big) \in \mathbb{C}^{M \times M}$ and $\bm{\Delta}_{\tilde{q}} \triangleq \text{diag}\big( \!\big[ e^{\jmath\tilde{\zeta}^{\tilde{q}}_{\tilde{1}}}, \dots, e^{\jmath\tilde{\zeta}^{\tilde{q}}_{\tilde{M}}} \big]\! \big) \in \mathbb{C}^{\tilde{M} \times \tilde{M}}$ are the phase-shift matrices of the $q$-th and $\tilde{q}$-th layers of the \ac{TX}- and \ac{RX}-\ac{SIM}, respectively, with $\zeta^{q}_{m}$ and $\tilde{\zeta}^{\tilde{q}}_{\tilde{m}}$ being the phase shifts of the $m$-th and $\tilde{m}$-th meta-atoms of the $q$-th and $\tilde{q}$-th layers, respectively.

In light of the above, the end-to-end \ac{SISO}\footnotemark\,\! time-varying channel, tunable via the configurations of the \ac{TX} and \ac{RX} \acp{SIM}, can be described by
\vspace{-1ex}
\begin{equation}
\label{eq:prop_SIM_channel}
\vspace{-1ex}
h(t,\tau) = \bm{u}\mathbf{R}_\mathrm{RX}^{1/2} \tilde{\mathbf{H}}(t,\tau) \mathbf{R}_\mathrm{TX}^{1/2}\bm{v},
\end{equation} 
where $\mathbf{R}_\mathrm{TX}$ and $\mathbf{R}_\mathrm{RX}$, respectively defined as $[\mathbf{R}_\mathrm{TX}]_{m,m'} \triangleq \text{sinc}\left(2\mathrm{d}_{m,m'}/\lambda\right)$ and $[\mathbf{R}_\mathrm{RX}]_{\tilde{m},\tilde{m}'} \triangleq \text{sinc}\big(2\tilde{\mathrm{d}}_{\tilde{m},\tilde{m}'}/\lambda\big)$, are \ac{TX} and \ac{RX} correlation matrices associated with the outermost layers of the corresponding \acp{SIM} \cite{AnJSAC2023}.
\vspace{-2ex}

\setcounter{footnote}{\thefootnote-2}
\footnotetext{This translates either to the assumption that the \acp{BS} are connected via a fronthaul network, or that the \ac{RX} \ac{BS} relies on pilot symbols.}

\setcounter{footnote}{\thefootnote+1}
\footnotetext{In \cite{KuranageTWC2024} algorithms to perform \ac{JCDE} as well as estimate delay and Doppler parameters in monostatic, multistatic and user-centric fashions in a \ac{SISO} setting were developed; in \cite{Ranasinghe_ICASSP_2024} an \ac{RPE} scheme to estimate delay, Doppler and \ac{AoA} in a \ac{MIMO} setting was presented; and \cite{RanasingheWCNC2025} a blind bistatic approach for the estimation of delay and Doppler was proposed. Unlike the system here discussed, however, in all such contributions the \ac{DD} channel model do not incorporate \acp{SIM}.}

\setcounter{footnote}{\thefootnote+1}
\footnotetext{We emphasize that a generalization to the \ac{MIMO} case requires only the extension of the \ac{SIM} transfer functions $\bm{v}$ and $\bm{u}$ to matrices, accordingly.}

In the case of a \ac{DD} channel, the $\tilde{M} \times M$ matrix $\tilde{\mathbf{H}}(t,\tau)$ in the \ac{SIM}-tunable model described in equation \eqref{eq:prop_SIM_channel} is given by\footnote{We highlight that a generalization of the work here presented to \ac{RIS}-aided scenarios can be achieved straightforwardly (if laboriously) by modifying $\tilde{\mathbf{H}}(t,\tau)$ accordingly, as shown in \cite{RanasingheDDSIM2025}.}
\vspace{-3ex}
\begin{eqnarray}
\label{eq:MIMO_TD_channel_general_SIM_LOS}
\tilde{\mathbf{H}}(t,\tau) \triangleq \sqrt{\tfrac{M\tilde{M}}{P}} \sum_{p=1}^P h_p e^{\jmath 2\pi \nu_p t} \delta\left(\tau-\tau_p\right)&&\\[-2ex] 
&&\hspace{-18ex}\times  \mathbf{b}_\mathrm{R}\left(\phi_p^{\rm in},\theta_p^{\rm in}\right) \mathbf{b}_\mathrm{T}\herm\left(\phi_p^{\rm out},\theta_p^{\rm out}\right),\nonumber
\vspace{-1ex}
\end{eqnarray}
with $\tau_p \in [0,\tau_\text{max}]$ and $\nu_p \in [-\nu_\text{max},\nu_\text{max}]$ denoting each $p$-th path's delay in seconds and Doppler shift in Hz, respectively; $\mathbf{b}_\mathrm{T}(\cdot,\cdot) \in \mathbb{C}^{M \times 1}$ and $\mathbf{b}_\mathrm{R}(\cdot,\cdot) \in \mathbb{C}^{\tilde{M} \times 1}$ respectively denoting the \ac{UPA} response vectors for the \ac{TX}-\ac{SIM} and \ac{RX}-\ac{SIM} as described in \cite{RanasingheDDSIM2025}; and $(\phi_p^{\rm in},\theta_p^{\rm in})$ and $(\phi_p^{\rm out},\theta_p^{\rm out})$ denoting the pairs of azimuth and elevation \acp{AoA} and \acp{AoD}, respectively, for each $p$-th signal propagation path having the complex channel gain $h_p$, with $p=\{1,\ldots,P\}$. 

The result of all the above is a \ac{MPDD} model, which in the sequel will be considered in determining the \ac{I/O} relationship between \ac{TX} and \ac{RX} signals, under the conditions resulting from the choice of a particular \ac{ISAC}-enabling waveform.

\vspace{-2ex}
\subsection{Signal Model under ISAC-Enabling Waveforms}
\label{sec:signal_model}

Let $\mathbf{x}$ be an $N\times 1$ vector of digitally-modulated information symbols to be transmitted over $N$ subcarriers using an \ac{ISAC}-enabling waveform which, for the sake of future convenience, will be assumed to be \ac{AFDM}.
It was shown in \cite{Rou_SPM_2024} that the corresponding received signal, after sampling, rectangular pulse-shaping, propagation through a multipath channel with $P$ paths, each with gain $\check{h}_p$, and demodulation, already taking into account the effect of introducing and removing cyclic prefixes, is given by
\vspace{-1ex}
\begin{equation*}
\mathbf{y} = \underbrace{\sum_{p=1}^P \check{h}_p \overbrace{
\mathbf{\Lambda}_2  \mathbf{F}_{N}  \mathbf{\Lambda}_1 \mathbf{\Theta}_p \mathbf{\Omega}^{f_p}  \mathbf{\Pi}^{\ell_p} \mathbf{\Lambda}_1\herm  \mathbf{F}_{N}\herm  \mathbf{\Lambda}_2\herm
}^{\triangleq\mathbf{G}^\text{AFDM}_p \in \mathbb{C}^{N \times N}}}_{\bar{\mathbf{H}} \in \mathbb{C}^{N \times N}}  \mathbf{x} + \bar{\mathbf{w}}
\in \mathbb{C}^{N \times 1},
\end{equation*}
\vspace{-6ex}
\begin{equation}
\quad
\label{eq:vectorized_TD_IO_kron}
\end{equation}
where $\mathbf{\Theta}_p \in \mathbb{C}^{N \times N}$, $\mathbf{\Omega}^{f_p} \in \mathbb{C}^{N \times N}$ and $\mathbf{\Pi}^{\ell_p} \in \mathbb{C}^{N \times N}$ are matrices that model\footnote{Due to space limitations, we omit several details such as the sampling rate, the normalization of delays and Doppler shits, etc.} the phase-shift effects of cyclic prefixes, Doppler shifts, and delay spreading, respectively \cite{Rou_SPM_2024}.

It was also shown in \cite{Rou_SPM_2024} that similar expressions also result if the information vector $\mathbf{x}$ is transmitted using other waveforms, sufficing to that end to modify the effective channel path matrix $\mathbf{G}_p$ accordingly.

For instance, in the case of \ac{OFDM} and \ac{OTFS}, instead of the matrix $\mathbf{G}_p$ implicitly defined in equation \eqref{eq:vectorized_TD_IO_kron} we would have respectively
\begin{subequations}
\label{eq:Gp_others}
\begin{eqnarray}
\label{eq:Gp_OFDM}
&\mathbf{G}^\text{OFDM}_p = \mathbf{F}_N  \mathbf{\Omega}^{f_p}  \mathbf{\Pi}^{\ell_p}\mathbf{F}_N\herm,&\\
\label{eq:Gp_OTFS}
&\mathbf{G}^\text{OTFS}_p = (\mathbf{F}_{N_1}\otimes\mathbf{I}_{N_2}) \mathbf{\Omega}^{f_p}  \mathbf{\Pi}^{\ell_p}  (\mathbf{F}_{N_1}\herm\otimes \mathbf{I}_{N_2}),&
\end{eqnarray}
where we highlight that for \ac{OFDM} and \ac{OTFS} $\mathbf{\Theta}_p=\mathbf{I}_N$ and the quantities $N_1$ and $N_2$ in the latter equation are both integers such that $N_1 \times N_2 = N$.
\end{subequations}

Finally, it was shown in \cite{RanasingheDDSIM2025} that with respect to the \ac{I/O} relationship described in equation \eqref{eq:vectorized_TD_IO_kron}, the parameterization features of channel model described in the Subsection \ref{sec:system_and_channel_model} affects only the channel path gains $\check{h}_p$, which in this case take the form
\vspace{-0.5ex}
\begin{equation}
\label{eq:check_H}
\!\!\!\check{h}_p \!\triangleq\! \sqrt{\tfrac{M \tilde{M}}{P}} h_p \bm{u}  \mathbf{R}_\mathrm{RX}^{1/2} \mathbf{b}_\mathrm{R}\left(\phi_p^{\rm in},\theta_p^{\rm in}\right) \mathbf{b}_\mathrm{T}\herm\left(\phi_p^{\rm out},\theta_p^{\rm out}\right) \mathbf{R}_\mathrm{TX}^{1/2}  \bm{v}.\!\!\!
\end{equation}

All in all, equations \eqref{eq:vectorized_TD_IO_kron} and \eqref{eq:check_H}, combined with \eqref{eq:Gp_others} when applicable, fully describe what \ac{AFDM}, \ac{OFDM} and \ac{OTFS} \ac{ISAC} signals are subjected to under a \ac{SISO} system aided by \ac{TX} and \ac{RX} \acp{SIM} in a \ac{DD} channel.

\vspace{-2ex}
\section{Proposed SIM-aided Bistatic ISAC Scheme}

\subsection{SIM Optimization for Bistatic Sensing}
\label{sec:sim_optimization}

Next, we consider the optimization of the bistatic \ac{ISAC} system described above, via the parameterization of the \ac{TX}/\ac{RX}-\acp{SIM}.
To that end, let us start by defining the quantities $\mathbf{B}_p \triangleq \mathbf{b}_\mathrm{R}\left(\phi_p^{\rm in},\theta_p^{\rm in}\right) \mathbf{b}_\mathrm{T}\herm\left(\phi_p^{\rm out},\theta_p^{\rm out}\right)$ and $\tilde{h}_p \triangleq h_p \sqrt{\tfrac{M \tilde{M}}{P}}$, such that we can formulate the following optimization problem
\begin{align}
\label{eq:path_optimization_sensing}
\underset{\bm{\mathcal{Z}},\tilde{\bm{\mathcal{Z}}}}{\max} \; \underset{p}{\min} & \underbrace{\big\|\tilde{h}_p \bm{u} \mathbf{R}_\mathrm{RX}^{1/2} \mathbf{B}_p \mathbf{R}_\mathrm{TX}^{1/2} \bm{v}\big\|^2}_{\triangleq\mathcal{O}(\bm{\mathcal{Z}},\tilde{\bm{\mathcal{Z}}})}\\
\text{\text{s}.\text{t}.} 
&\;|\zeta^{q}_{m}| \leq \pi \; \forall (q,m), \text{and\;} \,|\tilde{\zeta}^{\tilde{q}}_{\tilde{m}}| \leq \pi \; \forall (\tilde{q},\tilde{m}),\nonumber
\vspace{-2ex}
\end{align}
where $\bm{\mathcal{Z}}\triangleq\{\bm{\zeta}_1,\ldots,\bm{\zeta}_q,\ldots,\bm{\zeta}_Q\}$, with $\bm{\zeta}_q\triangleq[\zeta_1^q\ldots,\zeta_M^q]\trans$, and $\tilde{\bm{\mathcal{Z}}}\triangleq\{\tilde{\bm{\zeta}}_{\tilde{1}},\ldots,\tilde{\bm{\zeta}}_{\tilde{q}},\ldots,\tilde{\bm{\zeta}}_{\tilde{Q}}\}$, with $\tilde{\bm{\zeta}}_{\tilde{q}}\triangleq[\tilde{\zeta}_{\tilde{1}}^{\tilde{q}}\ldots,\tilde{\zeta}_{\tilde{M}}^{\tilde{q}}]\trans$.

\begin{subequations}
\label{eq:final_grad}
Although not convex, this problem can be solved efficiently via a subspace-wise gradient ascent algorithm, with the sub-gradients $\nabla_{\bm{\zeta}_{q}} \mathcal{O}(\bm{\mathcal{Z}},\tilde{\bm{\mathcal{Z}}})\in\mathbb{C}^{M\times 1}$ of the objective function in equation \eqref{eq:path_optimization_sensing} with respect to the vectors $\bm{\zeta}_q$ given by
\vspace{-1ex}
\begin{equation}
\label{eq:final_grad_TXSIM_sensing}
\!\!\!\!\!\!\!\!\!\nabla_{\!\bm{\zeta}_{q}} \mathcal{O}(\bm{\mathcal{Z}},\tilde{\bm{\mathcal{Z}}}) = 2 \Im \Big\{\mathbf{\Psi}_{q}\herm \tilde{\mathbf{f}}_{t:q,p}\herm \bm{\Upsilon}_{q,p} \bm{v} \Big\},
\end{equation}
with
\vspace{-1ex}
\begin{equation}
\bm{\Upsilon}_{q,p} \triangleq \big|\tilde{h}_p\big|^2 \mathbf{R}_\mathrm{TX}^{{\sf H}/2} \mathbf{B}_p\herm \mathbf{R}_\mathrm{RX}^{{\sf H}/2}  \bm{u}\herm \bm{u} \mathbf{R}_\mathrm{RX}^{1/2} \mathbf{B}_p \mathbf{R}_\mathrm{TX}^{1/2},
\vspace{-1ex}
\end{equation}
\vspace{-1ex}
\begin{equation}
\!\!\!\tilde{\mathbf{f}}_{t:q,p} \!\triangleq\!\! \prod_{q'=1}^{q+1}\!\! \bm{\Psi}_{Q-q'+1} \bm{\Gamma}_{Q-q'+1} \text{diag}\bigg(\bm{\Gamma}_{q}\!\!\!\!\!\!\! \!\prod_{q'= Q - q + 2}^{Q}\!\!\!\!\!\!\!\! \bm{\Psi}_{Q-q'+1} \bm{\Gamma}_{Q-q'+1}\!\bigg),\!\!\!
\end{equation}
and the sub-gradients $\nabla_{\!\tilde{\bm{\zeta}}_{\tilde{q}}} \mathcal{O}(\bm{\mathcal{Z}},\tilde{\bm{\mathcal{Z}}})\in\mathbb{C}^{\tilde{M}\times 1}$ with respect to the vectors $\tilde{\bm{\zeta}}_{\tilde{q}}$ obtained similarly.
\end{subequations}

With the closed-form expressions for the sub-gradients $\nabla_{\!\bm{\zeta}_{q}} \mathcal{O}(\bm{\mathcal{Z}},\tilde{\bm{\mathcal{Z}}})$ and $\nabla_{\!\tilde{\bm{\zeta}}_{\tilde{q}}} \mathcal{O}(\bm{\mathcal{Z}},\tilde{\bm{\mathcal{Z}}})$ as in equation \eqref{eq:final_grad}, problem \eqref{eq:path_optimization_sensing}
can be solved as follows.
First, using a random initialization of the \ac{TX}/\ac{RX} \acp{SIM} transfer functions $\bm{v}$ and $\bm{u}$, the path yielding the smallest value for the objective $\mathcal{O}(\bm{\mathcal{Z}},\tilde{\bm{\mathcal{Z}}})$ is computed.
Then, for that path, which determines the corresponding quantities $\tilde{h}_p$ and $\bm{B}_p$ in equation \eqref{eq:path_optimization_sensing}, the phases $\bm{\zeta}_{q}$ and $\tilde{\bm{\zeta}}_{\tilde{q}}$ determining the \acp{SIM} responses $\bm{v}$ and $\bm{u}$ are updated via\footnotemark

\begin{figure}[H]
  \centering
  {{\includegraphics[width=\columnwidth]{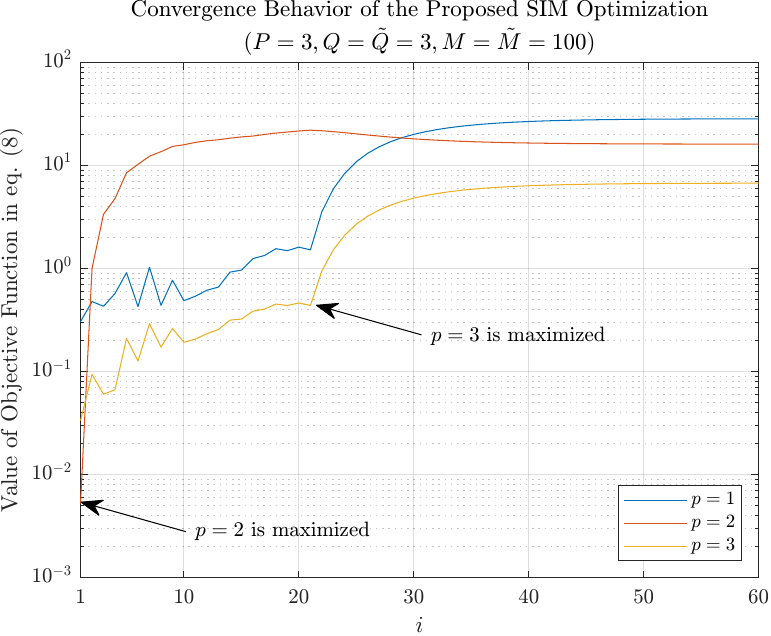}}}%
 \vspace{-2ex}
  \caption{The convergence behavior of the proposed SIM optimization in a typical case with $P = 3$ paths, $Q = \tilde{Q} = 3$ layers and  $M = \tilde{M} = 100$ meta-atoms per layer.}
  \label{fig:Conv_SIM_opt}
\end{figure}

\quad\\[-7ex]
\begin{subequations}
\label{eq:param_updates_GD}
\begin{equation}
\bm{\zeta}_{q}^{(i + 1)} = \bm{\zeta}_{q}^{(i)} + \lambda^{(i)} \vartheta^{(i)} \nabla_{\!\bm{\zeta}_{q}} \mathcal{O}(\bm{\mathcal{Z}},\tilde{\bm{\mathcal{Z}}}),
\vspace{-1ex}
\end{equation}
\begin{equation}
\tilde{\bm{\zeta}}_{\tilde{q}}^{(i + 1)} = \tilde{\bm{\zeta}}_{\tilde{q}}^{(i)} + \lambda^{(i)} \tilde{\vartheta}^{(i)} \nabla_{\!\tilde{\bm{\zeta}}_{\tilde{q}}} \mathcal{O}(\bm{\mathcal{Z}},\tilde{\bm{\mathcal{Z}}}),
\vspace{-1ex}
\end{equation}
\end{subequations}
where $\lambda^{(i)} \in (0,1)$ is the decaying learning rate parameter to ensure convergence, while $\vartheta^{(i)},\tilde{\vartheta}^{(i)}$ are normalization parameters calculated at each step as
\vspace{-1ex}
\begin{subequations}
\label{eq:normalization_params}
\begin{equation}
\vartheta^{(i)} = \pi / \underset{q, m}{\max}\nabla_{\!\bm{\zeta}_{q}} \mathcal{O}(\bm{\mathcal{Z}},\tilde{\bm{\mathcal{Z}}}),
\vspace{-1ex}
\end{equation}
\begin{equation}
\tilde{\vartheta}^{(i)} = \pi / \underset{\tilde{q}, \tilde{m}}{\max}\nabla_{\!\tilde{\bm{\zeta}}_{\tilde{q}}} \mathcal{O}(\bm{\mathcal{Z}},\tilde{\bm{\mathcal{Z}}}).
\vspace{-1ex}
\end{equation}
\end{subequations}

The convergence behavior of the optimization scheme here presented is illustrated in Fig. \ref{fig:Conv_SIM_opt}, in a setting with $P=3$.
It can be see in that example that initially the \ac{SIM} is optimized to bring out the second path ($p = 2$) and then, after a sufficient number of iterations, the focus moves to the third path ($p = 3$) which is the weakest path at after 20 iterations.
It can also be appreciated that subsequently, by optimizing for $p = 3$, the channel gains of all paths are improved, with the algorithm converging quickly at around 40 iterations.

\footnotetext{While $\bm{\zeta}_{q}$ and $\tilde{\bm{\zeta}}_{\tilde{q}}$ and the search for the path with weakest power can be updated after each iteration, for the purpose of complexity reduction these updates can also be done after a prescribed number of iterations are executed.}

\vspace{-1ex}
\section{Proposed Bistatic RPE Algorithm}
\label{sec:rpe}

\subsection{Radar Parameter Estimation Approach}
\label{sec:sim_optimization}

\begin{subequations}
\label{eq:sparse_RPE}
Since the worst case scenario of a \ac{DD} channel is characterized by pair of maximum normalized delay and Doppler shift, respectively $\max(\ell_p) \ll N$ and $\max(f_p) \ll N$, given a finite number $P$ of paths, the received signal described in equation \eqref{eq:vectorized_TD_IO_kron} can be recast as a double summation over both the normalized delay and Doppler taps as
\vspace{-0.5ex}
\begin{equation}
\label{eq:modified_input_output_relation_for_estimation}
\mathbf{y} \!=\!\!\! \sum_{\bar{k}=0}^{K_\tau - 1} \sum_{\bar{d}=0}^{D_\nu - 1} \hspace{-4ex}\underbrace{{\mathbf{G}_{\bar{k},\bar{d}}^\text{RPE}} \mathbf{x}}_{\qquad\triangleq\mathbf{e}_{\bar{k},\bar{d}}\in \mathbb{C}^{N\times 1}} \hspace{-4ex} h_{\bar{k},\bar{d}} + \bar{\mathbf{w}} = \mathbf{E} \mathbf{h} + \bar{\mathbf{w}} \in \mathbb{C}^{N\times 1},
\vspace{-1ex}
\end{equation}
where $K_\tau$ and $D_\nu$ are numbers large enough to define a fine grid discretizing the region determined by maximum normalized delay and Doppler shift, such that the only non-zero terms inside the summation are those for matching an actual path, that is, where both $\tau_{\bar{k}} = \tau_p$ and $\nu_{\bar{d}} = \nu_p$, and where the dictionary matrix $\mathbf{E} \in \mathbb{C}^{N\times K_\tau D_\nu}$ and the sparse channel vector $\mathbf{h} \in \mathbb{C}^{K_\tau D_\nu \times 1}$ are respectively defined as
\vspace{-1ex}
\begin{equation}
\label{eq:sparse_dict_matrix_def}
\mathbf{E}\! \triangleq\! [\mathbf{e}_{0,0}, \dots, \mathbf{e}_{0,D_\nu\! -\! 1}, \dots, \mathbf{e}_{K_\tau\! -\! 1,0}, \dots, \mathbf{e}_{K_\tau\! -\! 1,D_\nu \!-\! 1}],
\vspace{-0.5ex}
\end{equation}
\begin{equation}
\label{eq:sparse_channel_vector}
\mathbf{h} \!\triangleq\! [h_{0,0}, \dots, h_{0,D_\nu\! -\! 1}, \dots, h_{K_\tau\! -\! 1,0}, \dots, h_{K_\tau\! -\! 1,D_\nu\! -\! 1}]\trans.\!\!
\vspace{-1ex}
\end{equation}
\end{subequations}

Under such conditions, estimating the radar parameters $(\tau_p,\nu_p)\,\forall\,p$ amounts to estimating the $P$ channel gains such that $h_{\bar{k},\bar{d}}\neq 0$, such that the RPE problem can be formulated as a canonical sparse signal recovery problem.
In other words, the problem of estimating the set of radar parameters $\{\tau_p,\nu_p\}$, $\forall p$ reduces to estimating the sparse channel vector $\mathbf{h}$, with the delays and Doppler shifts obtained from the corresponding indices $(\bar{k},\bar{d})$ where $h_{\bar{k},\bar{d}}\neq 0$, given the received signal $\mathbf{y}$ and the dictionary matrix $\mathbf{E}$.

Notice that as a consequence of this approach, the assumption that $P$ is known is relaxed into the assumption that the channel paths are orthogonal in the delay-Doppler grid, such that $P$ can be unambiguously inferred from $\mathbf{h}$\footnote{For estimation purposes, we assume that while the number of \ac{UE} paths (hereafter denoted $P_U$) is known, the total number of paths $P$ is unknown.}.
We also point out that once the non-zero entries of $\mathbf{h}$ and the corresponding $\{\tau_p,\nu_p\}$ parameters are obtained, a discretized model of equation \eqref{eq:check_H} can also be build, such that the \acp{AoA} and \acp{AoD} associated with the distinct paths can also be estimated\footnote{In particular, for each known $\check{h}_p$, the estimation of the associated angles $\phi_p^{\rm in},\theta_p^{\rm in},\phi_p^{\rm out}$ and $\theta_p^{\rm out}$ can be formulated as an rank-constrained oblique manifold optimization problem or solved via a DoA estimation \cite{Ranasinghe_ICASSP_2024}.}.
We leave that problem, however, to be addressed in a follow-up work.

\vspace{-3ex}
\subsection{PDA-Based Radar Parameter Estimation Scheme}

In view of the sparse recovery-based \ac{RPE} approach described above, we are now ready to design the actual receiver to extract, based on equation \eqref{eq:sparse_RPE}, the radar parameters $\tau_p$ and $\nu_p,\,\forall\,p$.
To that end, let $\check{P} \triangleq K_\tau D_\nu$ be the total number of points in the grid, and $\check{p} = \{1,\dots,\check{P}\}$, and let us model the elements $\hat{h}_{\check{p}}$ of the sparse channel vector estimates as random variables following a Bernoulli-Gaussian type distribution, $i.e.$
\vspace{-1ex}
\begin{equation}
\label{eq:h_m_pdf}
\hat{h}_{\check{p}} \sim  p_{_{h_{\check{p}}}} \!(x; \kappa,\bar{h}_{\check{p}},\bar{\sigma}_{\check{p}}) \triangleq (1 - \kappa) \delta(x) + \kappa\,\mathcal{CN}\big(x;\bar{h}_{\check{p}},\bar{\sigma}_{\check{p}} \big) ,\!\!
\vspace{-1ex}
\end{equation}
where $\delta(x)$ is the centralized Dirac delta function and $\kappa$ is the sparsity of  $\mathbf{h}$ (updated as an index vector denoting the non-zero elements of $\mathbf{h}$), whose (unknown) true value is given by $\kappa = P/(K_\tau D_\nu)$, while $\bar{h}_{\check{p}}$ and $\bar{\sigma}_{\check{p}}$ are, respectively, the (also unknown) mean and variance of each variable, whose true value vary for each position $\check{p}$ in $\mathbf{\hat{\;h}}$, depending on whether a (and which) path matches the corresponding point in the grid.

Following steps similar to those \cite{KuranageTWC2024}, a \ac{PDA}-based message passing algorithm can be designed to estimate the aforementioned variables, which consists of the following steps.

\textbf{Soft Interference Cancellation:} The \ac{sIC} expression corresponding to an estimate of $h_{\check{p}}$, is given by
\vspace{-2ex}
\begin{equation}
\label{eq:Soft_IC_PDA}
\!\!\tilde{\mathbf{y}}_{{\check{p}}}^{(i)} \!=\! \mathbf{y} -\!\!\sum_{q\neq {\check{p}}} \!\mathbf{e}_{q} \hat{h}_{q}^{(i)}  \!=\! \mathbf{e}_{{\check{p}}}h_{\check{p}} \!+\!\!\!\! \overbrace{\sum_{q \neq {\check{p}}}^{K_\tau D_\nu}  \!\!(\mathbf{e}_{q} h_q \!-\! \mathbf{e}_{q} \hat{h}_{q}^{(i)}) \!+\! \tilde{\mathbf{w}}}^{\text{residual interference+noise component}}\!\!\!, 
\vspace{-1ex}
\end{equation}
where $\mathbf{e}_{\check{p}}$ is the ${\check{p}}$-th column of the dictionary matrix $\mathbf{E}$.
\newpage

It follows from the \ac{CLT} that, under large-system conditions, the residual interference-plus-noise component can be approximated by a multivariate complex Gaussian variate.
In other words, the \ac{VGA} can be applied such that the conditional \ac{PDF} of the beliefs $\tilde{\mathbf{y}}_{{\check{p}}}^{(i)}$, given $h_{\check{p}}$, can be expressed as
\vspace{-1ex}
\begin{equation}
\label{eq:VGA_y_given_h}
\tilde{\mathbf{y}}_{{\check{p}}}^{(i)} \sim p_{\textbf{y} | \text{h}_{{\check{p}}}} (\textbf{y} | h_{{\check{p}}})
\propto \text{exp} \big[ - \big( \textbf{y} - \mathbf{e}_{{\check{p}}} h_{{\check{p}}} \big)\herm \mathbf{\Sigma}^{-1(i)}_{{\check{p}}} \big( \textbf{y} - \mathbf{e}_{{\check{p}}} h_{{\check{p}}} \big) \big],
\end{equation}
where $\textbf{y}$ is an auxiliary variable, and the conditional covariance matrix $\mathbf{\Sigma}_{\check{p}}^{(i)}$, with $\sigma_w^2$ being the noise power, is given by
\vspace{-1ex}
\begin{align}
\mathbf{\Sigma}_{\check{p}}^{(i)} & \triangleq \mathbb{E}_{\textbf{h,$\tilde{\mathbf{w}}$}|\hat{h}_{\check{p}} \neq h_{\check{p}}} \Big[ \big( \tilde{\mathbf{y}}_{{\check{p}}}^{(i)} - \mathbf{e}_{{\check{p}}} h_{{\check{p}}} \big) \big( \tilde{\mathbf{y}}_{{\check{p}}}^{(i)} - \mathbf{e}_{{\check{p}}} h_{{\check{p}}} \big)\herm  \Big]
\nonumber \\[-1ex]
& = \sum_{q \neq {\check{p}}}^{K_\tau D_\nu} \hat{\sigma}_{h:q}^{2(i)} \mathbf{e}_q \mathbf{e}_q\herm + \sigma_w^2 \mathbf{I}_N.
\label{eq:covariance_matrix_PDA}
\vspace{-1ex}
\end{align}

\textbf{Belief Generation:} The beliefs associated with the estimate of the ${\check{p}}$-th channel entry $h_{\check{p}}$ can be obtained by combining the contributions of all \ac{sIC} beliefs $\tilde{\mathbf{y}}_{{\check{p}}}^{(i)}$, under the \ac{PDF}
\vspace{-1ex}
\begin{equation}
p_{\text{h} | h_{\check{p}}} (\text{h} | h_{\check{p}}) \propto \text{exp} \Big[ - \tfrac{|\text{h} - \tilde{h}_{\check{p}}^{(i)}|^2}{\tilde{\sigma}_{{\check{p}}}^{2(i)}} \Big],
\label{eq:ell_extrinsic_belief_PDA}
\vspace{-1ex}
\end{equation}
which yields
\vspace{-1ex}
\begin{subequations}
\label{eq:mean_and_var_extrinsic_belief_PDA}
\begin{equation}
\tilde{h}_{\check{p}}^{(i)} \triangleq \frac{1}{\eta_{\check{p}}^{(i)}} \mathbf{e}_{\check{p}}\herm \mathbf{\Sigma}^{-1(i)} \tilde{\mathbf{y}}_{{\check{p}}}^{(i)}\;\;\text{and}\;\;
\tilde{\sigma}_{{\check{p}}}^{2(i)} \triangleq \frac{1 - \eta_{\check{p}}^{(i)} \hat{\sigma}_{{\check{p}}}^{2(i)}}{\eta_{\check{p}}^{(i)}},
\vspace{-1ex}
\end{equation}
where $\eta_{\check{p}}^{(i)}$ is a normalization factor defined as
\vspace{-1ex}
\begin{equation}
\label{eq:eta_PDA}
\eta_{\check{p}}^{(i)} \triangleq \mathbf{e}_{\check{p}}\herm \mathbf{\Sigma}^{-1(i)} \mathbf{e}_{\check{p}},
\vspace{-1ex}
\end{equation}
and the common conditional covariance matrix\footnote{The matrix inversion lemma is used in the derivation of \eqref{eq:mean_and_var_extrinsic_belief_PDA}, such that the same inverse matrix $\mathbf{\Sigma}^{(i)}$ can be used instead of $\mathbf{\Sigma}_{\check{p}}^{(i)}$.}
is given by
\vspace{-1ex}
\begin{equation}
\mathbf{\Sigma}^{(i)} \triangleq \sum_{{\check{p}}=1}^{K_\tau D_\nu} \hat{\sigma}_{{\check{p}}}^{2(i)} \mathbf{e}_{\check{p}} \mathbf{e}_{\check{p}}\herm + \sigma_w^2 \mathbf{I}_N.
\vspace{-1ex}
\end{equation}
\end{subequations}

\textbf{Soft Replica Generation:} Under the \ac{VGA}, the soft replicas of $h_{\check{p}}$ can be inferred from the conditional expectation given the extrinsic beliefs and the fact that the effective noise components in $\hat{h}_{\check{p}}^{(i)}, \forall {\check{p}}$ are uncorrelated.
Subsequently, leveraging the assumption that $h_{\check{p}}$ follows a Bernoulli-Gaussian distribution and using the Gaussian-\ac{PDF} multiplication rule \cite{Parker_TSP_2014}, the denoising equations can be expressed as
\vspace{-1ex}
\begin{subequations}
\begin{equation}
\label{eq:BG_update_sparsity_rate}
\hat{\kappa}_{{\check{p}}}^{(i)} \!\triangleq\! \Bigg(\! \frac{1 - {\kappa}}{{\kappa}}  \frac{\tilde{\sigma}_{{\check{p}}}^{2(i)} + \hat{\sigma}_{{\check{p}}}^{2(i-1)}}{\tilde{\sigma}_{{\check{p}}}^{2(i)}} \: e^{- \frac{|\tilde{h}_{{\check{p}}}^{(i)}|^2}{\tilde{\sigma}_{{\check{p}}}^{2(i)}} + \frac{|\tilde{h}_{{\check{p}}}^{(i)} - \hat{h}_{{\check{p}}}^{(i-1)}|^2}{\tilde{\sigma}_{{\check{p}}}^{2(i)} + \hat{\sigma}_{{\check{p}}}^{2(i-1)}}} \!+ 1 \!\Bigg)^{\!\!-1}\!\!\!\!,
\vspace{-1ex}
\end{equation}
\vspace{-1ex}
\begin{equation}
\label{eq:BG_update_rules_h}
\hat{h}_{{\check{p}}}^{(i)} \triangleq \frac{\hat{\sigma}_{{\check{p}}}^{2(i-1)} \tilde{h}_{{\check{p}}}^{(i)} + \tilde{\sigma}_{{\check{p}}}^{2(i)} \hat{h}_{{\check{p}}}^{(i-1)}}{\tilde{\sigma}_{{\check{p}}}^{2(i)} + \hat{\sigma}_{{\check{p}}}^{2(i-1)}},
\vspace{-1ex}
\end{equation}
\begin{equation}
  \label{eq:BG_update_rules_sigma}
  \hat{\sigma}_{{\check{p}}}^{2(i)} \triangleq \frac{\hat{\sigma}_{{\check{p}}}^{2(i-1)} \tilde{\sigma}_{{\check{p}}}^{2(i)}}{\tilde{\sigma}_{{\check{p}}}^{2(i)} + \hat{\sigma}_{{\check{p}}}^{2(i-1)}},
  \vspace{-1ex}
  \end{equation}
\end{subequations}
where the unknown true parameters $\bar{h}_{\check{p}}$ and $\bar{\sigma}_{\check{p}}$ defined in equation \eqref{eq:h_m_pdf} are replaced by the prior estimates $\hat{h}_{\check{p}}^{(i-1)}$ and $\hat{\sigma}_{{\check{p}}}^{2(i-1)}$, respectively\footnote{The unknown parameters can also be updated via an \ac{EM} algorithm.
However, this addition is relegated to a follow-up work.}.

\newpage

\begin{algorithm}[H]
\caption{SIM Optimization and RPE for Bistatic ISAC}
\label{alg:proposed_estimator}
%
\setlength{\baselineskip}{11pt}
\textbf{Input:} Receive signal vector $\mathbf{y}$, dictionary matrix $\mathbf{E}$, number of \ac{UE} paths $P_U$, average channel power per path $\sigma_h^2$, damping factor $\tilde{\beta}_h$, number of \ac{PDA} iterations $i_{\max}$, number of gradient descent iterations $i_{\mathrm{GD}}$, and noise variance $\sigma^2_w$. \\
\textbf{Output:} Estimates $\hat{\tau}_p$ and $\hat{\nu}_p$ extracted from the non-zero indices of the sparse channel estimate vector $\!\hat{\;\mathbf{h}}$.
\vspace{-2ex} 
\begin{algorithmic}[1]  
\STATEx \hspace{-3.5ex}\hrulefill
\STATEx \hspace{-3.5ex}\textbf{Initialization}
\STATEx \hspace{-3.5ex} - Set counter to $i = 0$ and ${\kappa} = P_U/(K_\tau D_\nu)$.
\STATEx \hspace{-3.5ex} - Set average channel power per path as $\sigma_h^2 = 1/(K_\tau D_\nu)$.
\STATEx \hspace{-3.5ex} - Set initial estimates $\hat{h}_{{\check{p}}}^{(0)} = 0$ and ${\hat{\sigma}}_{{\check{p}}}^{(0)} = \sigma_h^2$ $,\forall {\check{p}}$.
\STATEx \hspace{-3.5ex} - Set paths\footnotemark as $\hat{P} = P_U$.
\STATEx \hspace{-3.5ex}\hrulefill
\STATEx \hspace{-3.5ex}\textbf{SIM Optimization via Greedy Steepest Ascent}
\STATEx \hspace{-3.5ex}\textbf{for} $\hat{p}=1$ to $\hat{P}+1$
\STATE Compute $\hat{p}$ for which \eqref{eq:path_optimization_sensing} is minimized.
\STATEx \textbf{for} $i=1$ to $i_{\mathrm{GD}}$ \textbf{do} $\forall q, \tilde{q}, m, \tilde{m}$
\STATE \hspace{3.5ex} Compute the gradients using~\eqref{eq:final_grad}.
\STATE \hspace{3.5ex} Update normalization parameters from~\eqref{eq:normalization_params}.
\STATE \hspace{3.5ex} Update the phase parameters via~\eqref{eq:param_updates_GD}.

\STATEx \textbf{end for}
\STATEx \hspace{-3.5ex}\textbf{end for}
\STATEx \hspace{-3.5ex}\textbf{Radar Parameter Estimation via PDA}
\STATEx \hspace{-3.5ex}\textbf{for} $i=1$ to $i_\text{max}$ \textbf{do} $\forall {\check{p}}$
\STATE Compute soft signal vectors $\tilde{\mathbf{y}}_{{\check{p}}}^{(i)}$ from equation \eqref{eq:Soft_IC_PDA}.
\STATE Compute beliefs $\tilde{h}_{\check{p}}^{(i)}$ and their
variances $\tilde{\sigma}_{{\check{p}}}^{2(i)}$ from \eqref{eq:mean_and_var_extrinsic_belief_PDA}.
\STATE Denoise indices $\hat{\kappa}_{\check{p}}^{(i)}$, estimates $\hat{h}_{{\check{p}}}^{(i)}$  and variances $\hat{\sigma}_{{\check{p}}}^{2(i)}$ from~\eqref{eq:BG_update_sparsity_rate}, \eqref{eq:BG_update_rules_h} and \eqref{eq:BG_update_rules_sigma}, respectively.
\STATE Damp estimates $\hat{h}_{{\check{p}}}^{(i)}$ and variances $\hat{\sigma}_{{\check{p}}}^{2(i)}$ using~\eqref{eq:soft_rep_and_MSE_updates}.
\STATEx \hspace{-3.5ex}{\textbf{end}} {\textbf{for}} 
\STATE Compute the estimates $\hat{\tau}_p$ and $\hat{\nu}_p$ corresponding to the indices ${\check{p}}$ of the non-zero entries of $\!\hat{\;\mathbf{h}}$ in accordance to expression~\eqref{eq:sparse_channel_vector}.
\end{algorithmic}
\end{algorithm}
\vspace{-2ex}

\footnotetext{In principle, $\hat{P}$ can also be updated after each execution of the \ac{PDA} algorithm in an iterative fashion, but this extension is relegated to a journal version of the paper.}

Finally, the soft replica $\hat{h}_{{\check{p}}}^{(i)}$ and its \ac{MSE} $\hat{\sigma}_{{\check{p}}}^{2(i)}$ can be, in general, obtained from the conditional expectation as\footnote{Note the already incorporated damping procedure to prevent convergence to local minima due to incorrect hard-decision replicas.}
\vspace{-1ex}
\begin{subequations}
\label{eq:soft_rep_and_MSE_updates}
\begin{equation}
\label{eq:PDA_soft_rep_update}
\hat{h}_{\check{p}}^{(i)} = \tilde{\beta}_h \hat{\kappa}_{{\check{p}}}^{(i)} \hat{h}_{{\check{p}}}^{(i)} + (1 - \tilde{\beta}_h) \hat{h}_{\check{p}}^{(i-1)},
\end{equation}
\begin{equation}
\label{eq:PDA_MSE_update}
\hat{\sigma}_{{\check{p}}}^{2(i)} = \tilde{\beta}_h \big[(1 - \hat{\kappa}_{{\check{p}}}^{(i)}) \hat{\kappa}_{{\check{p}}}^{(i)} |\hat{h}_{{\check{p}}}^{(i)}|^2 + \hat{\kappa}_{{\check{p}}}^{(i)} \hat{\sigma}_{{\check{p}}}^{2(i)}\big] + (1-\tilde{\beta}_h) \big[ \hat{\sigma}_{h:{\check{p}}}^{2(i-1)} \big].
\end{equation}
\end{subequations}

The complete pseudocode for the \ac{SIM} parameterization and sensing procedure proposed is summarized in Algorithm \ref{alg:proposed_estimator}.



\vspace{-1ex}
\section{Numerical Results}
\label{sec:numerical_results}

\subsection{Sensing Performance Results}
\label{subsec:sensing_perf_analy}
\begin{figure}[H]
\subfigure[{\footnotesize MSE Performance}]%
{\includegraphics[width=\columnwidth]{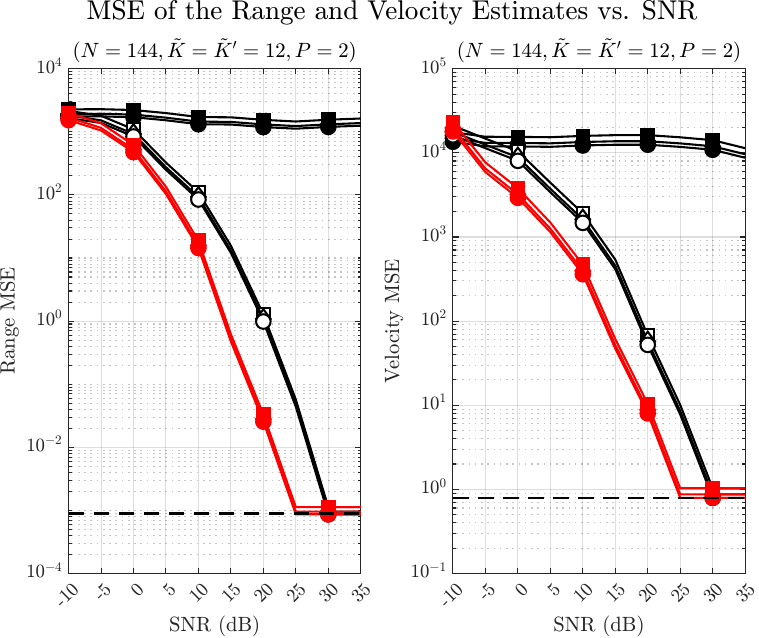}
\label{fig:MSE_SIM_SISO}}\\
\subfigure[{\footnotesize BER Performance}]%
{\includegraphics[width=\columnwidth]{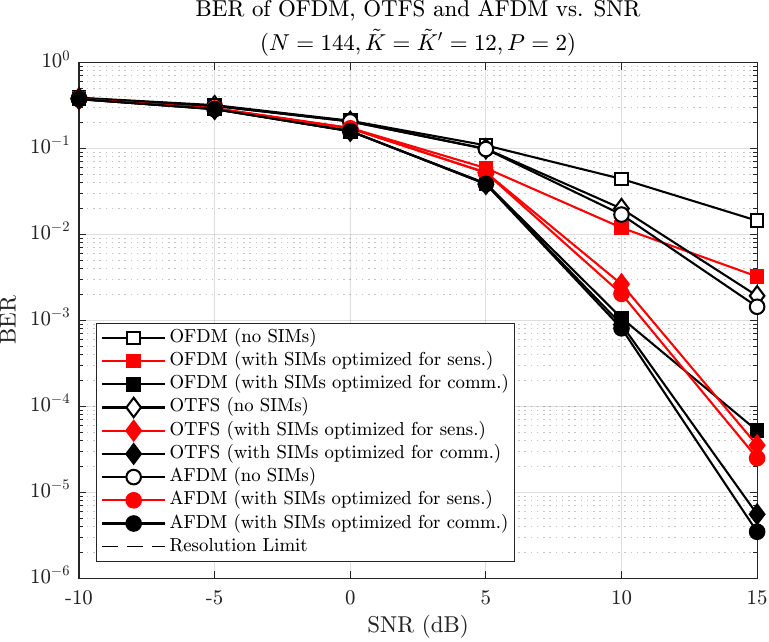}
\label{fig:BER_SIM_SISO_ISAC}}
\vspace{-2ex}
\caption{\ac{ISAC} performance of \ac{OFDM}, \ac{OTFS} and \ac{AFDM} waveforms with \ac{QPSK} modulation in \ac{MPDD} channels with high-mobility, with \ac{SIM} placed at very close distances to both the \ac{TX} and the \ac{RX} in a \ac{SISO} setting.}
\label{fig:BER_MSE_Alg1}
\vspace{-1ex}
\end{figure}

Fig.~\ref{fig:MSE_SIM_SISO} illustrates the \ac{MSE} performance of the aforementioned method with \ac{OFDM}, \ac{OTFS}, and \ac{AFDM} signaling\footnote{It was shown in \cite{GaudioTWC2020,KuranageTWC2024} that the communication-centric \ac{ISAC} performance for \ac{OFDM}, \ac{OTFS}, and \ac{AFDM} are very similar, which holds true even with the \ac{SIM}.} with \ac{QPSK} modulation in a \ac{SIM}-enabled \ac{DD} channel, considering a \ac{SISO}
setting where both the \ac{TX} and \ac{RX} are equipped with \ac{SIM}.
\newpage

It is assumed that the system operates with a carrier frequency of $28$ GHz ($i.e.$, at wavelength $\lambda = 10.7$ mm) with a bandwidth of $B = 20$ MHz in an environment with identical \ac{TX} and \ac{RX} \ac{SIM}, both with $Q = \tilde{Q} = 3$ layers of metasurfaces and $M = \tilde{M} = 100$ meta-atoms per layer, leading to $M_x = M_z = \tilde{M}_x = \tilde{M}_z = 10$.
Finally, the sampling frequency is set to $F_\mathrm{S}=B$ and the number of symbols per frame to $N = 144$.
For the \ac{MPDD} channel, the \ac{2D} and \ac{3D} elevation \acp{AoD}/\acp{AoA} are uniformly distributed in $[0,\pi]$, while the \ac{3D} azimuth \acp{AoD}/\acp{AoA} are in $[-\frac{\pi}{2},\frac{\pi}{2}]$.

Within the channel, there are two targets ($i.e., P = P_U = 2$), with ranges $37.5$ m and $97.5$ m and velocities $-54$ m/s and $+54$ m/s, respectively. 
In Fig.~\ref{fig:MSE_SIM_SISO}, the resolution limit defines how fine the search grid is, i.e., it measures the estimation accuracy for the range and velocity if the estimate for $\!\hat{\;\mathbf{h}}$, obtained as the output of the proposed Algorithm~\ref{alg:proposed_estimator}, is perfect.
As observed in the figure, there exists a large gain in performance when using the sensing-based optimization for the \ac{SIM}, as opposed to \ac{RPE} with no \acp{SIM} while the communication-based optimization presented in \cite{RanasingheDDSIM2025} fails completely.

\vspace{-2ex}
\subsection{Communications Performance Results}

A natural question that occurs is whether there is a fundamental tradeoff between the performance of the communication and sensing functionalities due to the worse performance of the \ac{SIM} optimized for communications in a sensing setting.

Under the same parameters as in the previous Section~\ref{subsec:sensing_perf_analy}, the communications performance of the \ac{SIM} optimized for sensing is also investigated.
As seen from Fig.~\ref{fig:BER_SIM_SISO_ISAC}, while the communications performance decreases from the case where the \ac{SIM} is optimized for communications in \cite{RanasingheDDSIM2025}, there is still a significant gain over the case where no \ac{SIM} are used, highlighting the effectiveness of the \ac{SIM} for \ac{ISAC}, regardless of the waveform in use.
However, it is noteworthy that the \ac{OFDM} waveform exhibits a smaller performance enhancement as opposed to \ac{OTFS} and \ac{AFDM} when using the sensing-optimized \acp{SIM}, hinting at the usual superiority of \ac{OTFS} and \ac{AFDM} waveforms in DD channels \cite{KuranageTWC2024}. 

\vspace{-2ex}
\section{Conclusion}
\label{sec:conclusions}

We proposed a \ac{SIM} optimization and \ac{RPE} schemes to improve the performance bistatic \ac{ISAC} under \ac{DD} channels.
The \ac{SIM} optimization approach seeks to maximize the channel gain of the weakest path, and was solved via a gradient ascent algorithm with closed-form expressions, while the \ac{RPE} scheme was designed based on a \ac{PDA} approach enabled by a novel sparse-recovery formulation of the sensing problem suitable to \ac{DD} channels.
Numerical results show a significant gain in both sensing and communication functionalities due to the impact of the emerging \ac{SIM} technology.



\vspace{-2ex}

\end{document}